\documentclass[reprint,superscriptaddress,aps,prb,twocolumn]{revtex4-2}
\usepackage{setspace}
\usepackage{amsmath}
\usepackage{graphicx}
\usepackage{amsfonts}
\usepackage{amssymb}
\usepackage{epstopdf} 
\usepackage{xcolor}
\usepackage{verbatim}
\usepackage{soul}
\usepackage{comment}
\usepackage{placeins}
\usepackage{textcomp}
\usepackage{bm}
\usepackage[colorlinks=true,bookmarks=false,citecolor=blue,urlcolor=blue]{hyperref}

\DeclareGraphicsExtensions{.pdf,.eps,.png,.jpg,.mps} 
\usepackage{float}

\begin{document}

\title{High-coherence hybrid-integrated 780 nm source by self-injection-locked second-harmonic generation in a high-Q silicon-nitride resonator}

\author{Bohan Li$^{1\ast}$, Zhiquan Yuan$^{1\ast}$, Warren Jin$^{3\ast}$, Lue Wu$^{1}$, Joel Guo$^{2}$, Qing-Xin Ji$^{1}$,Avi Feshali$^{3}$,Mario Paniccia$^{3}$,John E. Bowers$^{2\dagger}$,Kerry J. Vahala$^{1\dagger}$ \\ 
$^1$T. J. Watson Laboratory of Applied Physics, California Institute of Technology, Pasadena, CA, USA \\
$^2$ECE Department, University of California Santa Barbara, Santa Barbara, CA, USA \\
$^3$Anello Photonics, Santa Clara, CA, USA\\
$^\ast$These authors contributed equally to this work. \\ 
$^\dagger$vahala@caltech.edu,jbowers@ucsb.edu}

\begin{abstract}
By self-injection-locking a 1560 nm distributed feedback semiconductor laser to a high-$Q$ silicon nitride resonator,  a high-coherence 780 nm second harmonic signal is generated via the photogalvanic-induced second-order nonlinearity. A record-low frequency noise floor of 4 Hz$^2$/Hz is achieved for the 780 nm emission. The approach can be generalized for signal generation over a wide range of visible and near-visible bands.
\end{abstract}

\maketitle
Highly coherent visible laser sources play a crucial role in the operation of optical atomic clocks \cite{ludlow2015optical}, automotive LiDAR \cite{hecht2018lidar}, and sensing systems \cite{degen2017quantum}. However, existing bench-top visible lasers are both costly and bulky, limiting their use beyond laboratory environments including application in future navigation and sensing systems. To address this challenge, we generate visible light in a high-Q silicon nitride microcavity that is hybridly-integrated to a semiconductor laser operating in the near-infrared band. The cavity both line narrows the laser through self-injection-locking (SIL) \cite{jin2021hertz,li2021reaching} and generates the high-coherence visible signal as a second-harmonic (SH) signal by way of the photogalvanic field-induced second-order nonlinearity \cite{lu2021efficient} and the all-optical-poling effect \cite{billat2017large,nitiss2022optically} in Si$_3$N$_4$. Frequency noise is reduced by 100-fold compared with previous integrated visible lasers \cite{siddharth2022near,corato2023widely,ling2023self,franken2021hybrid}. The approach can be readily tuned to any visible or near-visible band. 

\begin{figure*}[t!]
\centering\includegraphics[width=\linewidth]{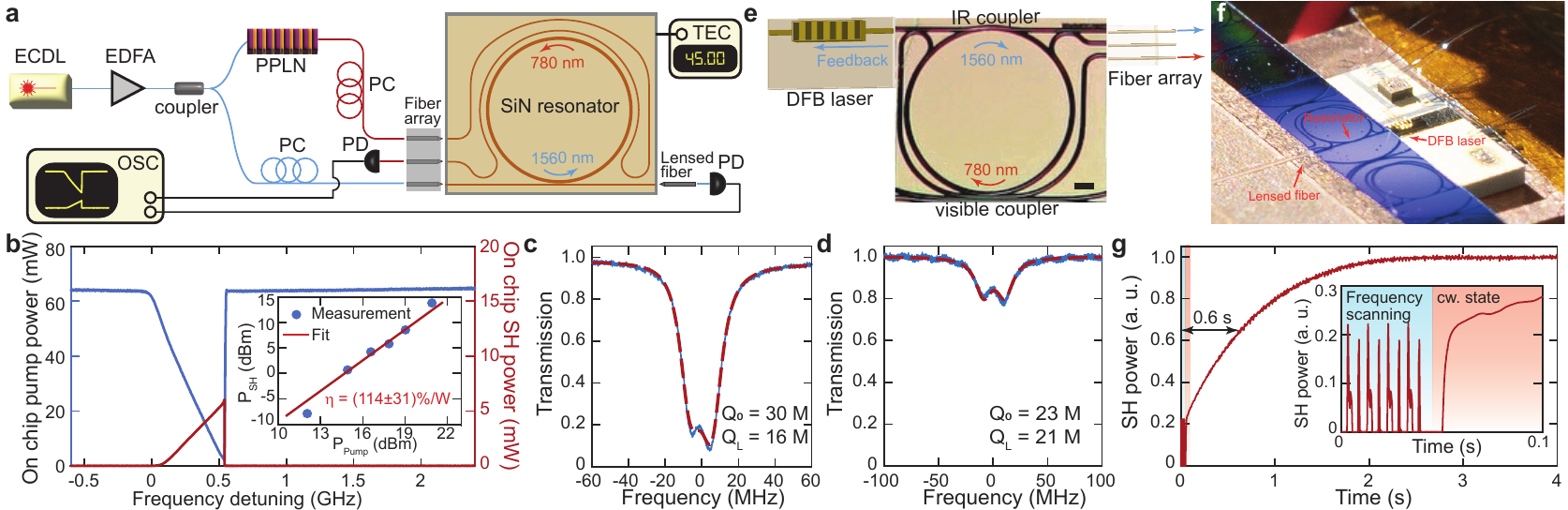}
\caption{Experimental setups and characterization of the Si$_3$N$_4$ resonator and SH generation performance.
(a) Experimental setup used to determine the phase matching condition and characterize the SH performance of the Si$_3$N$_4$ resonator. ECDL, external cavity diode lasers; EDFA, Erbium-doped fiber amplifier; PPLN, periodically poled lithium niobate; PC, polarization controller; PD, photodiode; OSC, oscilloscope; TEC, thermoelectric cooler.
(b) Measured on-chip transmission pump power (left axis) and generated SH power (right axis) when scanning the frequency of a tunable laser across a cavity resonance at the phase matching condition.
Inset: measured on-chip SH power (blue dots) plotted versus pump power levels. The frequency conversion efficiency is fit by the red line.
(c, d) Transmission spectra of the pump resonance at 1560.1 nm (c) and the SH resonance at 780.05 nm (d). The experimental measurements are plotted in blue and the theoretical fittings are plotted in red. Each resonance exhibits backscatter-induced splitting.
(e) Schematic of the hybrid-integrated frequency conversion laser, where a DFB laser is endfire-coupled to a high-Q  silicon-nitride
microresonator to provide feedback to the laser. The upper and lower waveguides are designed for coupling at 1560 nm and 780 nm, respectively. The scale bar is 200 $\mu$m.
(f) Photograph of the hybrid-integrated frequency conversion laser in panel (e).
(g) The time response of the SH power after the DFB laser frequency is stopped at the operation point.
Inset: zoom-in of the main plot. Prior to fixing the DFB laser frequency, the pump laser frequency is scaanned and SH power changes periodically during frequency scanning (blue region). When the scanning is stopped, SH power builds to steady state (red region).
}
\label{fig1}
\end{figure*}
The resonator is fabricated using the ultra-low-loss silicon-nitride photonic platform \cite{jin2021hertz,puckett2021} with a 100 nm thick silicon nitride waveguide core and a 2.2 $\mu$m thick silica top cladding. The resonator has a 5 $\mu$m waveguide width and a 850 $\mu$m radius. Two pulley couplers are designed for efficient coupling at both the near-infrared and visible bands. The near-infrared coupler has a 2.3 $\mu$m waveguide width with a 3.5 $\mu$m gap that is designed to prevent coupling to the visible mode. The visible coupler has a 1.6 $\mu$m waveguide width with a 0.3 $\mu$m gap that is designed to reduce coupling to the near-infrared mode.

The resonator is first characterized using the experimental setup shown in Fig. \ref{fig1}a. A 4-channel fiber array and a lensed fiber are used to couple to near-infrared and visible resonances simultaneously. To probe the resonances, the output of a near-infrared tunable laser is split with one output doubled in frequency using a periodically-poled lithium niobate (PPLN) crystal. In this way, first and second harmonic probe waves are generated to characterize resonator spectra in these bands.  Because the photogalvanic effect induces optical poling to thereby establish the quasi-phase-matching condition \cite{nitiss2022optically}, a visible mode having twice the pumping frequency will achieve SH generation regardless of its propagation constant. To establish this condition, the Si$_3$N$_4$ chip is temperature controlled to tune the mode spectra. Tuning over no more than one free-spectral-range is sufficient, and in the current setup a chip temperature of 45 $^\circ$C aligns the pump resonance at 1560.1 nm with a visible resonance at 780.05 nm. The $Q$ factors of these two modes are determined by transmission spectra measurements shown in Fig. \ref{fig1}c,d. Photogalvanic-induced second-harmonic generation can be readily observed when scanning the pump laser across the near-infrared resonance, as shown in Fig. \ref{fig1}b. Continuous-wave SH power measurements with the pump laser frequency fixed at the cavity resonance are shown in the inset of Fig. \ref{fig1}b. The SH conversion efficiency is estimated to be 114$\pm$31\%W (average over measured powers) and SH output power as high as 24 mW is measured (in the waveguide). 

To achieve a high coherence visible light source in a compact foot print, we replace the bulk tunable laser with a distributed-feedback (DFB) chip laser as shown in the experimental setup in Fig. \ref{fig1}e,f. The DFB laser is endfire-coupled to the silicon-nitride chip and can deliver 20 mW pump power to the resonator waveguide (accounting for 6 dB facet coupling loss). Backscatter-induced feedback from the resonator to the laser provides self-injection-locking that dramatically reduces the laser frequency noise \cite{jin2021hertz,li2021reaching}. Upon current tuning the DFB laser frequency into the 1560.1 nm resonance, the optical poling process is initialized through the field-induced photogalvanic effect. This process can be monitored by scanning the frequency of the DFB laser around the resonance by modulating the pump current with a function generator. The SH signal produced during forward and backward scanning is shown in the inset of Fig. \ref{fig1}g (blue region). The function generator is then turned off, and the DFB frequency self-locks into the resonance center. The resulting SH signal time evolution is shown over short time interval in the inset of Fig. \ref{fig1}g (red region) and over several seconds in the main panel. The SH power build-up features a 0.6s rise time and takes only a few seconds to reach steady state. In steady-state operation, the SH power at 780 nm reaches over 0.5 mW on-chip as monitored via a lensed fiber (12dB coupling loss). In comparison to minutes-level build-up times in silicon nitride waveguides \cite{billat2017large}, this relatively short build-up time is attributed to cavity enhancement \cite{lu2021efficient}. 

\begin{figure}[t!]
\centering\includegraphics[width=\linewidth]{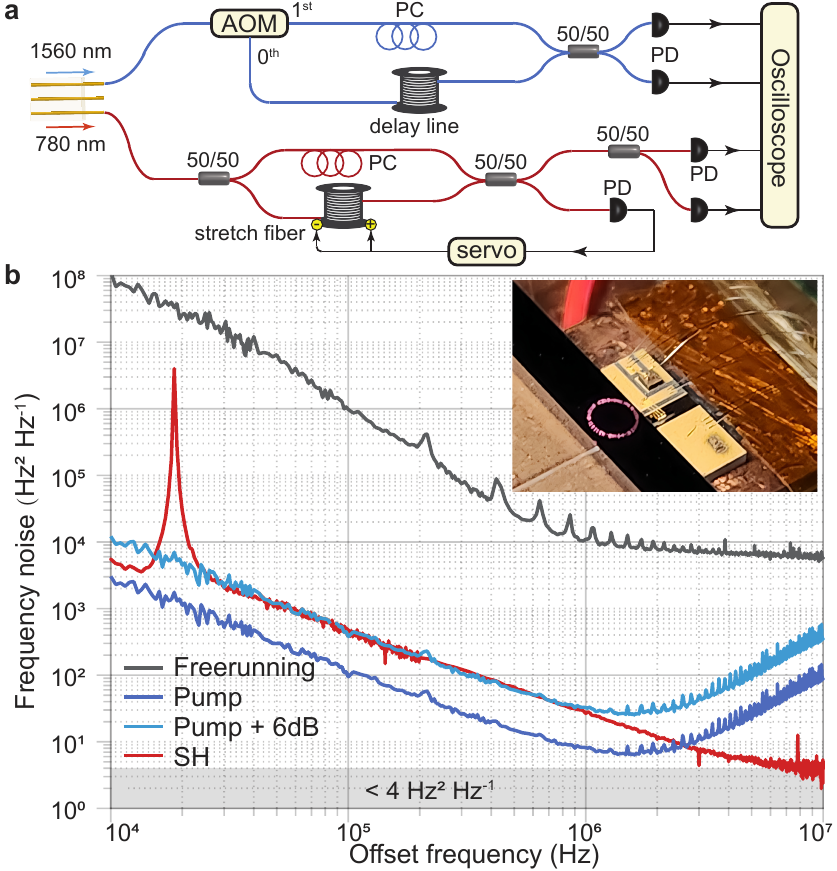}
\caption{Frequency noise measurement for SIL pump laser and generated SH laser.
(a) Frequency noise measurement setups for both the SIL 1560nm laser and the SH 780 nm laser. AOM, acousto-optic modulator. 
(b) Measured single-sideband frequency noise spectrum. Gray, blue and red traces show the frequency noise spectra of the free-running DFB signal, SIL pump signal (1560 nm), and SH signal (780 nm), respectively. The light blue trace shows the SIL pump laser frequency noise up-shifted by 6 dB. The inset is a photograph of the device under operation with 780 nm emission visible on the ring resonator.}
\label{fig2}
\end{figure}

The SIL 1560 nm light and SH generated 780 nm light are then further analyzed to determine their frequency noise performance (Fig. \ref{fig2}a). The SH signal is sent to a delayed self-homodyne detection setup with quadrature-point locking \cite{lee2012chemically} and its measured frequency noise is shown as the red trace in Fig. \ref{fig2}b. The peak at 18 kHz offset in the spectrum is due to the feedback loop response of a fiber stretcher used to maintain the quadrature point. At high offset frequencies, the photodetector (PD) white noise is suppressed using a cross-correlation technique \cite{walls1992cross, yuan2022correlated}, and achieves 4 Hz$^2$/Hz noise floor above 6 MHz offset frequency, corresponding to a record-low 25 Hz instantaneous linewidth for visible on-chip sources. The frequency noise of the self-injection locked pump laser is characterized with a self-heterodyne approach \cite{yuan2022correlated}, and the result is shown as the blue curve in Fig. \ref{fig2}. Compared with the free-running DFB laser noise (gray trace), the SIL process suppresses the noise by 40 (32) dB at 100 kHz (1 MHz) offset frequency. The generated SH laser noise is 4 times higher than the SIL pump laser noise due to coherent photon conversion \cite{ling2023self}. The difference in these spectra at high offset frequencies is due to spontaneous emission noise in the 1560 nm signal resulting from its measurement at the transmission port of the coupled resonator system. 

\begin{figure}[t!]
\centering\includegraphics[width=\linewidth]{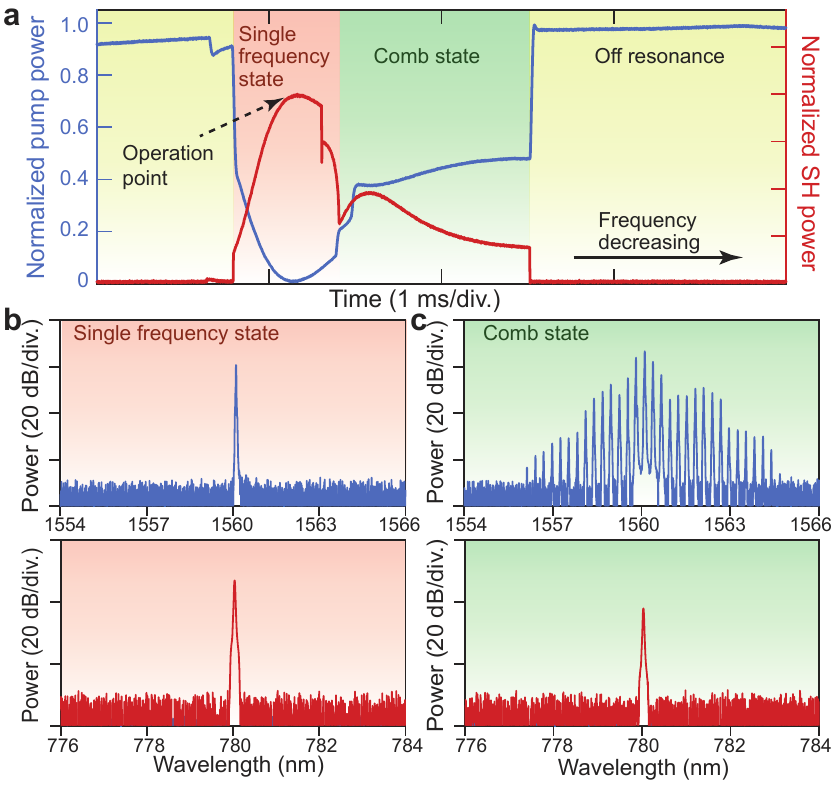}
\caption{Self-injection locked laser response with pump current induced frequency tuning.
(a) The measured SIL pump power (blue) and second harmonic power (red) when the pump current is scanned across the resonance. When the DFB laser frequency is tuned into the cavity resonance, the near infrared laser is initially in a single frequency state and eventually tunes to a comb state.
(b, c) Upper panels: Measured optical spectra of the near infrared laser in the single frequency state (b) and comb state (c). Lower panels: corresponding SH 780 nm spectra.
}
\label{fig3}
\end{figure}

High intra-cavity photon density and resonant backscattering make this system prone to Kerr frequency comb generation \cite{jin2021hertz}. Though important in many applications, frequency comb generation needs to be avoided here as it diverts power in the pump mode away from the SH generation process. SIL comb formation is governed by feedback phase and frequency detuning \cite{shen2020integrated,lihachev2022platicon}, and these parameters also provide a way to favor single mode lasing over comb generation. In the present device the laser-to-chip gap (feedback phase) and pump current (frequency detuning) provide useful controls. The latter is illustrated in Fig. \ref{fig3}a where the transmitted pumping power and the SH power are plotted versus DFB laser current scan. Distinct regimes where the single-frequency SIL state and the comb state appear are indicated. In the comb state, only the pump comb line can be frequency doubled due to phase matching condition so that the generated SH power is reduced. The operation point used in the previous power and noise measurements is indicated. Typical SH spectra in the two regimes are shown in Fig. \ref{fig3}b,c. 

In conclusion, we have demonstrated a record-low 4 Hz$^2$/Hz frequency noise floor for a hybrid-integrated visible light source by self-injection-locking a DFB laser with a high-$Q$ Si$_3$N$_4$ resonator. This device can be readily heterogeneously integrated upon slight revision of the coupling waveguides \cite{xiang2023three}. The current resonator intrinsic $Q$ factor is not optimal, and based on prior work could exceed 250M at 1560nm \cite{jin2021hertz}. This would further increase SH power and lower frequency noise levels since the SIL noise reduction scales as $Q^2$ \cite{kondratiev2017self}. Moreover, this scaling makes generation of highly coherent signals easier in the near-IR where optical Qs are overall much higher. The SH-SIL process extends this advantage into the visible bands. Finally, the photogalvanic effect makes access to other wavelengths straightforward. Devices require only waveguide couplers designed for efficient visible and near infrared operation.  

During preparation of this paper two other papers were reported: one on direct generation of high coherence visible light \cite{zhang2023photonic}  and one on SIL-SH using the photogalvanic effect \cite{bresCLEO}.

\vbox{}
\noindent \textbf{\large Acknowledgements} \\  
This project was supported by the Defense Advanced Research Projects Agency (DARPA) through LUMOS (grant no. HR001-20-2-0044).

\vbox{}
\noindent \textbf{\large Competing interests} \\ 
The authors declare no competing interests.

\vbox{}
\noindent \textbf{\large Data Availability} \\ The data that support the plots within this paper and other findings of this study are available from the corresponding authors upon reasonable request.

\bibliography{main.bib}

\end{document}